# The participation of public in knowledge production: a citizen science projects overview

Núria Bautista-Puig[*], Enrique Orduña-Malea[**] and Philippe Mongeon[***]

[*]*nuriabau@ucm.es*
https://orcid.org/0000-0003-2404-0683
Library and Information Science Department, Complutense University of Madrid, Spain

[**]*enorma@upv.es*
https://orcid.org/0000-0002-1989-8477
Department of Audiovisual Communication, Documentation and History of Art, Universitat Politècnica de València, Spain

[***]*PMongeon@dal.ca*
https://orcid.org/0000-0003-1021-059X
Department of Information Science, Dalhousie University, Canada

Citizen Science (CS) is related to public engagement in scientific research. The tasks in which the citizens can be involved are diverse and can range from data collection and tagging images to participation in the planning and research design. However, little is known about the involvement degree of the citizens to CS projects, and the contribution of those projects to the advancement of knowledge (e.g. scientific outcomes). This study aims to gain a better understanding by analysing the SciStarter database. A total of 2,346 CS projects were identified, mainly from Ecology and Environmental Sciences. Of these projects, 91% show low participation of the citizens (Level 1 'citizens as sensors' and 2 'citizens as interpreters', from Haklay's scale). In terms of scientific output, 918 papers indexed in the Web of Science (WoS) were identified. The most prolific projects were found to have lower levels of citizen involvement, specifically at Levels 1 and 2.

## 1. Introduction

Recognizing that global challenges (e.g. climate change, COVID, Agenda 2030) are complex and cannot be managed solely through a scientific approach, science is moving to new participatory ways of generating knowledge (Monzón-Alvarado et al., 2020) that integrate different perspectives and forms of knowledge that can help to better understand our complex socio-environmental systems, improve decision-making, and strengthen democracies. As this perspective can be a *win-win* situation for both sides (Marzuki, 2015), public participation initiatives have gained wider political, institutional and public attention (e.g. 'science with and for society' funding in H2020 programme).

Citizen science (CS hereafter) is one of the many forms of public engagement in scientific knowledge production in which citizens are directly involved in different ways, such as intellectual, knowledge, tools or resources (Socientize Project, 2013). The most important benefits of CS include increasing public participation (e.g. scientific literacy), time - and cost – effectiveness, and increased geographic reach (e.g. reaching remote locations) (Conrad & Hilchey, 2011; West & Pateman, 2017). However, the uncertainty regarding data collection, data use and data quality remains a scientific concern (Ali et al., 2019; Conrad & Hilchey, 2011). Other unquiets identified in the literature are those related to ethical issues and the invisibility of volunteer contributors (Elliott & Rosenberg, 2019; Kullenberg & Kasperowski, 2016). The European Citizen Science Association (ECSA) proposed a set of ten principles to address these challenges, one of which is to acknowledge the contribution of citizens in the research outputs.

Although CS has existed for a long time, it has expanded in recent years (i.e. publications have increased by 2200%[1] over the 1995-2018 period (Pelacho et al., 2020)). This is also evidenced by the steadily increasing number of new CS projects registered on SciStarter, a citizen science hub where project leaders can register and describe their projects)[2]. This rise has been partially attributed to an increase in the public awareness (Fritz, 2019; Quinlivan et al., 2020) and to the spread of emerging technologies (mobile devices, social media, low-cost sensors and cloud computing) that have facilitated data collection, processing and communication among volunteers (Socientize Project, 2013).

Citizen science has been used in diverse fields such as health (Belansky et al., 2011), environment (Conrad & Hilchey, 2011; Sullivan et al., 2014), astronomy (Jordan Raddick et al., 2013) and social sciences (Tauginienė et al., 2020). Ecological and environmental sciences have led the charge for CS with prominent efforts, showing some of the longest-running projects that have contributed meaningful data to science and conservation (Dickinson et al., 2010; Kosmala et al., 2016). For example, the eBird Project (https://ebird.org), launched in 2002, generated more than 200 publications.

The level and nature of the citizens' involvement in the research process can vary from one CS project to other. Citizens can be involved in tasks ranging from data collection and tagging images to research planning and design. Different participation scales have been proposed to capture the degree of intensity of citizen involvement. The first was Arnstein's (1969) 'ladder of citizen participation', ranging from 'non-participation' to 'citizen power'. More recently, Haklay (2013) proposed a four-level scale: from citizens used as sensors (Level 1) to the full engagement of volunteers (Level 4). While increased integration of the citizens in the research is possible and potentially desirable, previous studies suggest that most CS projects are characterized by a low participation level (Schleicher & Schmidt, 2020).

From a bibliometric perspective, previous studies have analysed the CS research output (Bautista-Puig et al., 2019; Kullenberg & Kasperowski, 2016; Pelacho et al., 2020). However, to the best of our knowledge, no studies have addressed the nature and extent of citizens' contribution to CS projects. In addition, the contribution of CS projects to specific scientific outcomes (i.e. publications) has not been covered so far. The aim of this study is thus to gain a better understanding of the citizen involvement (participation degree), and the scientific outcomes of the CS projects, by answering the following research questions:

1. How are CS projects distributed across fields?
2. What is the citizen involvement (in terms of degree) of these CS projects?
3. What scientific outcomes have been generated from the CS projects? What topics do they cover?

---

[2] Data extracted from the projects (around 200 in 2011) available at: https://blog.scistarter.org/2015/10/ten-citizen-science-projects-and-events-added-to-scistarter-this-week/. The number of active registered projects is 1,508 at the time of writing this paper

## 2. Data and Methods

This project uses data from the SciStarter database, which is supported by the Arizona State University and constitutes "an online community dedicated to improving the citizen science experience for project managers and participants". Data extraction using web scraping techniques was carried out in March 2023. For each project, the following information was collected: name of the project, task, goal, description, and field (see sample in Figure 1). Fields are classified into 28 different categories (e.g. Agriculture, Biology), and projects could be multi-classified. For instance, 'Journey North', which tracks wildlife migration and season change, is linked to both Biology and Astronomy. Data were obtained in JSON format and subsequently transformed into CSV files. After a cleaning procedure (e.g. duplicates and dissemination events were removed), the process yielded 2,346 projects. Although some projects seem similar (e.g. the National Parks BioBlitz project implemented in different regions), the tasks and focus of the projects are quite diverse.

Figure 1. Screenshot of the Galaxy Zoo project's description sheet
Source: scistarter.org

The involvement degree of citizen in CS projects was obtained by using the Haklay classification (2013). To do this, the projects were semi-automatically classified in the following four levels: 1) Crowdsourcing (citizens as sensors); 2) Distributed Intelligence (Citizens as Interpreters); 3) Participatory Science (Participation in problem detection and data collection); 4) Extreme Citizen Science (Collaboration in problem definition, data collection and analysis). The coding of each project was based on the description of the tasks provided by Scistarter (e.g. in the goal, task and description fields). For instance, a project in which citizens carry a sensor and the cognitive engagement is minimal, it would be considered 'Crowdsourcing' (level 1). In contrast, participation in research design, data collection and analysis would be considered 'Extreme Citizen Science' (level 4).

Otherwise, several search strategies were designed to collect data on the research outcomes by using the title of the projects within the acknowledgment fields (FO, FG and FD field tags) in Web of Science (WoS) Core Collection (SCI, SSCI, A&HCI). The query for some projects was combined with TS=citizen science to reduce false positives (e.g. BEACH). Data were downloaded in March 2023. After a cleaning process, only one-eighth of the data was considered, resulting in a final set of 918 unique documents (all document types were considered). Most of these projects do not have any identifier, except projects like EC-funded (e.g. 'Doing It Together Science'). This reflects the difficulty in identifying outcomes from these projects.

CorTexT Manager software (Breucker et al., 2016) was used to identify the publication topics and their evolution over time. Single and multi-terms (n-grams) were extracted from each title, abstract and keywords of those papers based on a trade-off between their specificity (using the chi2score) and frequency.

## 3. Findings

*3.1. Distribution of projects by field*
2,346 projects were collected from SciStarter database. Of these, 1652 projects (70,4%) are assigned to one field category at least. 'Ecology & Environment' (1035; 44.12%), 'Nature & Outdoors' (928; 29.56%), 'Biology' (785; 33.46%) and 'Animals' (604; 25.75%) stand outs in the number of projects (Figure 2). This concurs with previous studies (e.g. Schleicher & Schmidt, 2020) that show the relevance of ecological and environmental sciences. A plausible explanation is that in these fields it is easier to design a CS project (e.g. sighting of animals) rather than in other fields (e.g. Law). Worth mentioning is the high number of projects in social sciences (113). However, this category groups a wide range of projects that involves human interactions in different fields (e.g. eBirds, Atlas of Living Australia, Cloudy with a Change of Pain).

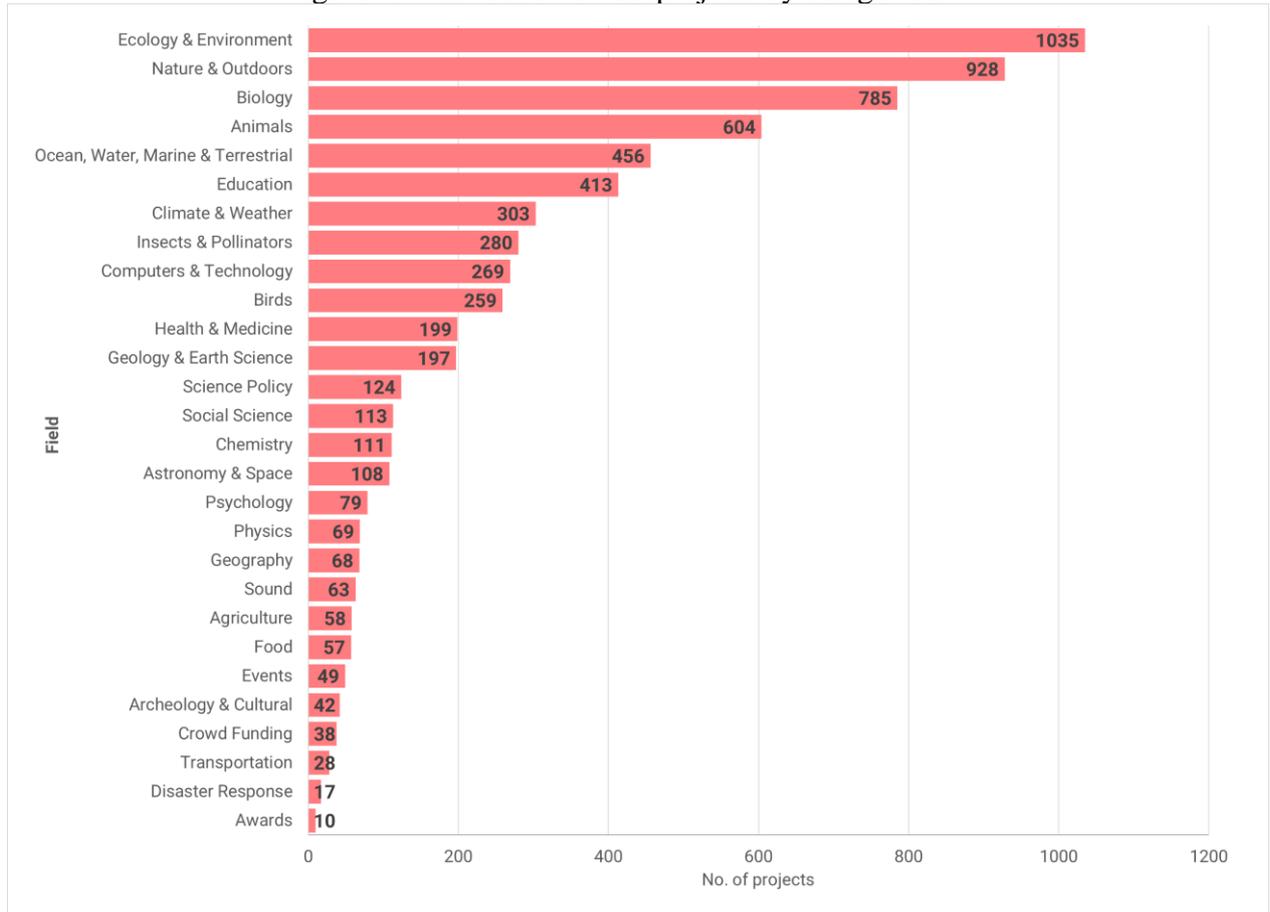

Figure 2. Distribution of CS projects by categories.

*3.2. Citizen involvement degree*

90.1% of the projects present a low citizen involvement (40% of projects in Level 1 and 51% of Level 2). Projects with higher involvement of citizens are scarce: 148 projects (6.31%) in Level 3; and 67 projects (2.8%) in Level 4.

Figure 3 shows the distribution of involvement degree by field. 'Archaeology & Cultural', 'Crowdfunding', 'Transportation', 'Chemistry', 'Food', 'Physics' and 'Astronomy' show the higher involvement of citizens within their projects in Level 4 (>12% of the projects). However, projects about 'Animals', 'Insects & Pollinators', 'Geography', 'Nature & Outdoors' and 'Birds' present the lowest participation (>40% of the projects). This denotes that the classic CS method, where participants help and involve with simple but cognitive activities (e.g. counting birds, sightings of animals) is still representative of the majority of projects in the sample. Nonetheless, SciStarter database presents some limitations that should be stressed. For example, 9.3% of the projects do not include the goal information; 11% the tasks, and 0.8% a description of the project.

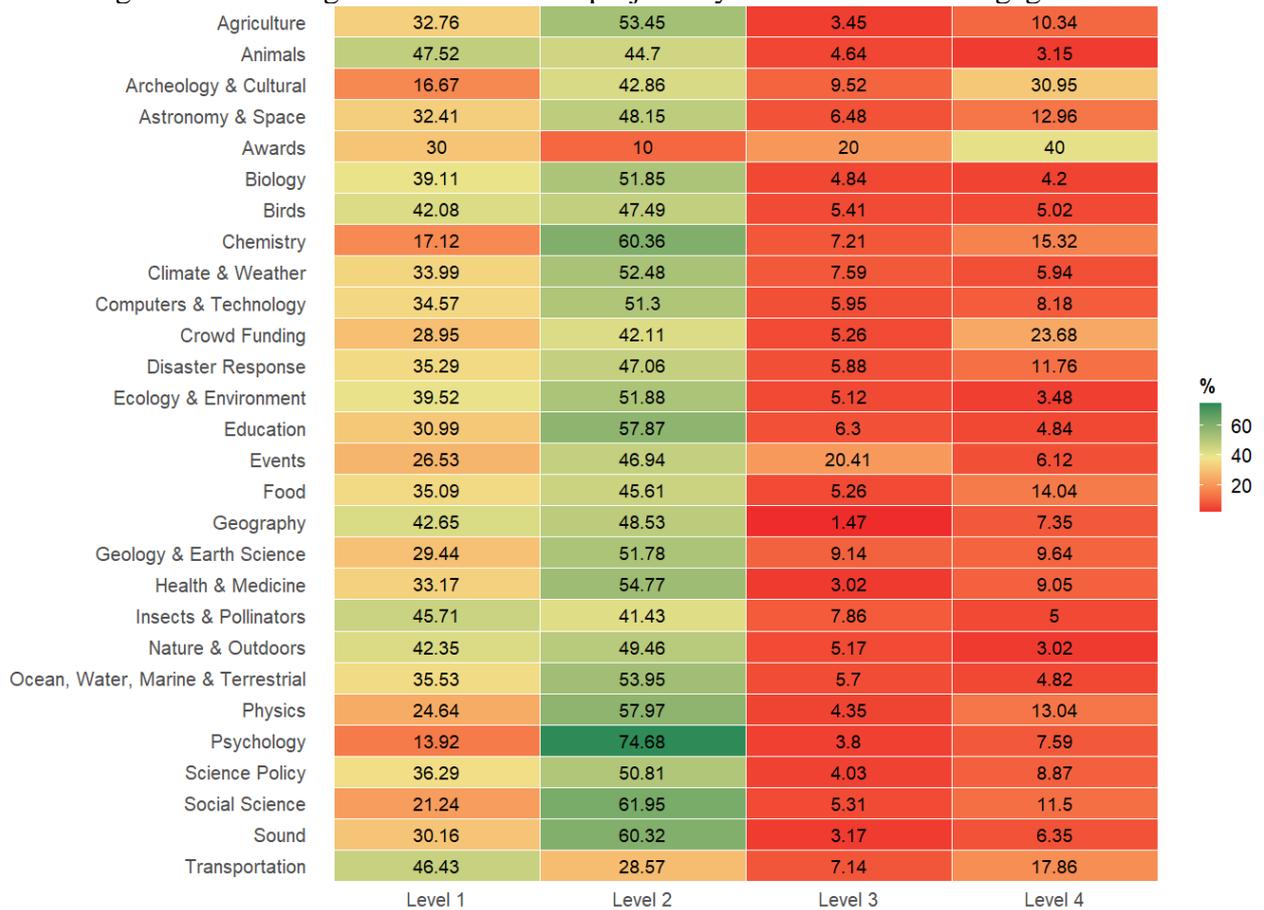

Figure 3. Percentage of distribution of projects by field and level of engagement

### 3.3. Scientific outputs from CS projects

918 papers mentioning any of the CS projects have been identified in WoS. The analysis revealed the most prolific projects are related to Biology (for example, 'Encyclopedia of Life' with 121 papers, 'Biodiversity Monitoring' with 94, and 'Save the Tasmanian Devil' with 79) and astronomy (DISCOVER-AQ with 44, and Galaxy Zoo with 34). Table 1 includes the distribution of the number of projects and publications from those projects, as well as the average level of participation by field. As we can observe, those projects labelled in fields that presented a higher output (e.g. Ecology & Environment, Nature, Animals, or Biology) show the lowest participation (>Level 2). A plausible explanation is that scientists fear that a co-design process could reduce the scientific quality, and that it requires an investment of time to the citizens (e.g. to train the citizens) (Schleicher & Schmidt, 2020).

Table 1. Distribution of projects, outcomes and level of participation by field

| Field | No. Projects | No. Papers | Level of participation (avg) |
|---|---|---|---|
| Ecology & Environment | 1035 | 540 | 1.80 |
| Nature & Outdoors | 928 | 485 | 1.82 |
| Animals | 604 | 402 | 1.65 |
| Biology | 785 | 363 | 1.81 |
| Ocean, Water, Marine & Terrestrial | 456 | 306 | 1.87 |
| Education | 413 | 300 | 1.91 |
| Birds | 259 | 229 | 1.88 |
| Insects & Pollinators | 280 | 204 | 1.90 |

| Topic | | | |
|---|---|---|---|
| Computers & Technology | 269 | 186 | 1.83 |
| Health & Medicine | 199 | 153 | 1.78 |
| Transportation | 28 | 106 | 2.00 |
| Astronomy & Space | 108 | 93 | 2.08 |
| Climate & Weather | 303 | 81 | 1.93 |
| Geology & Earth Science | 197 | 73 | 2.10 |
| Social Science | 113 | 67 | 2.00 |
| Physics | 69 | 60 | 1.83 |
| Science Policy | 124 | 57 | 1.78 |
| Chemistry | 111 | 56 | 2.18 |
| Psychology | 79 | 55 | 2.00 |
| Archeology & Cultural | 42 | 28 | 2.50 |
| Crowd Funding | 38 | 26 | 2.33 |
| Sound | 63 | 22 | 2.25 |
| Agriculture | 58 | 15 | 2.29 |
| Geography | 68 | 15 | 1.67 |
| Awards | 10 | 12 | 2.50 |
| Food | 57 | 7 | 2.25 |
| Events | 49 | 6 | 2.50 |
| Disaster Response | 17 | 4 | 1.00 |

Figure 4 shows the distribution of the publication topics over time. Overall, the results shown a predominance in environmental (e.g. with the discovery of new species or specific regions) and astronomy (e.g. spiral galaxies, brightness) topics over time. There are two tubes about CS data and data quality. The first (2009-2013) is related to environmental projects (e.g. bird monitoring and air quality monitoring) whereas the second (2009-2022) stresses the role of the individuals (citizen scientists) and the quality of those data in diverse fields (e.g. biodiversity, water quality).

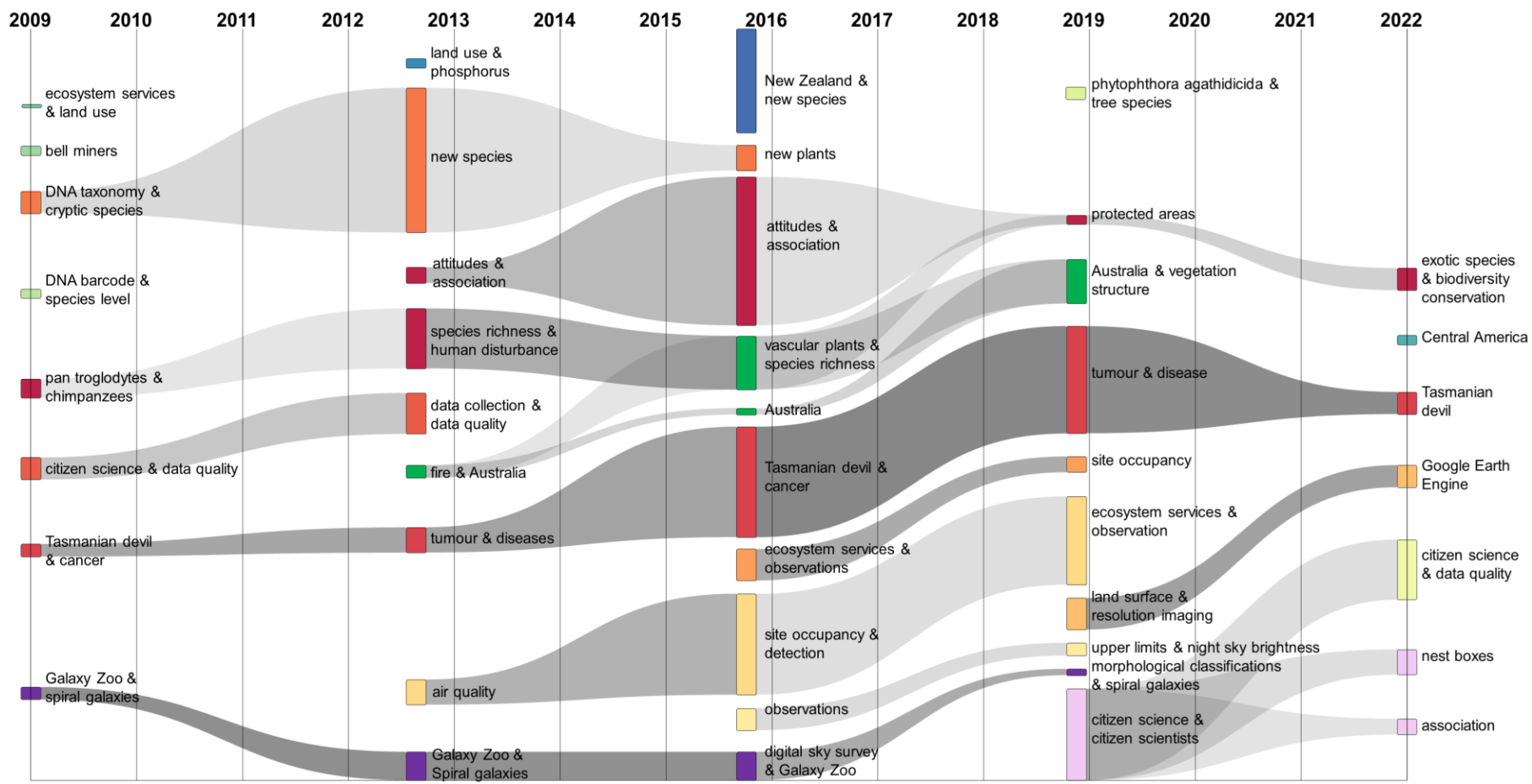

Figure 4. Sankey diagram of the evolution of topics over time

## 4. Conclusions

The preliminary results obtained have contributed to better understanding of the knowledge base on CS by providing a descriptive analysis of the CS project fields, participation degree, and scientific outcomes of those projects. This way, this exploratory study evidence that (i) the involvement degree of citizens is limited (only 2.8% of the projects are Level 4); (ii) CS projects are mainly environmental-focused; (iii) although the scientific output is scarce (918 papers indexed in WoS), the projects that present a higher output (>30 papers) have the lowest participation (>Level 2).

However, results should be taken cautiously. First, all citizen science projects are not covered by the SciStarter database, which also presents a bias toward Western countries. Second, the classification of the levels of citizen involvement depends on the description of the tasks, which might be inaccurate. Third, considering only the WoS database in the study might introduce limitations due to the underrepresentation of other related published works (e.g. reports, working papers), which may be relevant to this specific topic. Fourth, the use of the publication titles could not capture the whole picture of the research when identifying the topics covered by the publications mentioning CS projects.

Future development of this work should include additional data sources for CS projects (e.g. eu-citizen.science, Zooniverse database) and research outputs (e.g. Dimensions, Google Scholar, Dimensions). In addition, the study will be complemented by employing qualitative research to determine the motivations from a top-down perspective (researchers) and bottom-up (e.g. the citizens to participate in those projects). This will provide solid foundations for further research and new avenues for fostering the broader adoption of CS initiatives.